# Software Engineering Practices for Scientific Software Development: A Systematic Mapping Study

*Elvira-Maria Arvanitou[1], Apostolos Ampatzoglou[1], Alexander Chatzigeorgiou[1], Jeffrey C. Carver[2]*

[1] Department of Applied Informatics, University of Macedonia, Thessaloniki, Greece

[2] Department of Computer Science, University of Alabama

e.arvanitou@uom.edu.gr, a.ampatzoglou@uom.edu.gr, achat@uom.gr, carver@cs.ua.edu

***Background***: The development of scientific software applications is far from trivial, due to the constant increase in the necessary complexity of these applications, their increasing size, and their need for intensive maintenance and reuse. ***Aim***: To this end, developers of scientific software (who usually lack a formal computer science background) need to use appropriate software engineering (SE) practices. This paper describes the results of a systematic mapping study on the use of SE for scientific application development and their impact on software quality. ***Method***: To achieve this goal we have performed a systematic mapping study on 359 papers. We first describe a catalogue of SE practices used in scientific software development. Then, we discuss the quality attributes of interest that drive the application of these practices, as well as tentative side-effects of applying the practices on qualities. ***Results***: The main findings indicate that scientific software developers are focusing on practices that improve implementation productivity, such as code reuse, use of third-party libraries, and the application of "*good*" programming techniques. In addition, apart from the finding that performance is a key-driver for many of these applications, scientific software developers also find maintainability and productivity to be important. ***Conclusions***: The results of the study are compared to existing literature, are interpreted under a software engineering prism, and various implications for researchers and practitioners are provided. One of the key findings of the study, which is considered as important for driving future research endeavors is the lack of evidence on the trade-offs that need to be made when applying a software practice, i.e., negative (indirect) effects on other quality attributes.

## 1. Introduction

Scientific software development refers to the analysis, design, implementation, testing, and deployment of software applications for scientific purposes (e.g., physics, biology, medical analysis, and data science). The need for continuous experimentation and validation of techniques (e.g., simulations and cases studies) before the release of scientific results has led to the emergence of the domain of scientific software development as an important method for researchers to be successful in multiple fields (Birdsall and Langdon, 1991). As a result, "*many scientists and engineers spend much of their lives writing, debugging, and maintaining software, but only a handful have ever been taught how to do this effectively: after a couple of introductory courses, they are left to rediscover (or reinvent) the rest of programming on their own. The result? Most spend far too much time wrestling with software, instead of doing research, but have no idea how reliable or efficient their programs are.*" (Wilson 2006). While this quote is 15 years old, the sentiment has not changed. If anything, the dependence upon software has increased within the scientific domain while scientists are still not well-equipped.

The US National Science Foundation (NSF) has awarded more than US$9.6 billion to support more than 18,000 projects and 95% of postdocs surveyed report the use software (that in most of the cases they have developed themselves) to support their research (Nangia and Katz, 2017). In addition, a recent blog post describing the results of a survey of 1,200 researchers funded by the US NSF showed that the vast majority of respondents did not have sufficient time for training and that most development activities (other than coding) were not well-supported by current development tools (Carver, 2019). Although scientists invest a large fraction of their time (more than 40%) to building software, they often do not take full advantage of the advancements in software engineering (SE) (Heaton and Carver, 2015). This lack of SE practices can be attributed, at least partially, to limited knowledge of the benefits of these practices (Schmidberger and Brügge, 2012). Only about half of the postdocs from the survey mentioned earlier had received any software development training (Nangia and Katz, 2017) and 75% of NSF-funded researchers reported no time for training (Carver, 2019). As one specific example, only about half of scientists know the basics of testing (Wilson 2006).

**Figure 1:** State of Practice and Envisioned Practices in Scientific Software Development

Based on the findings above, the left side of Figure 1 highlights some issues that arise when scientific software developers lack proper SE practices. The text below explains these issues in more detail:

- ***Management of Large Code-bases and Collaboration***. Scientific software applications can be complex, often containing millions of lines of code (Méndez et al., 20104). Projects of this complexity cannot be developed by a single person. Therefore, scientific software developers need to use collaborative software development approaches and tools. In addition, projects of this scale are often multi-disciplinary (Howison and Herbsleb, 2011), which also increases the need for collaboration. For example, development of a full-scale application can require input from scientists with different expertise (e.g., mathematicians, biologists, natural scientists, etc.).

- *Maintainability*. New development is only a portion of the software lifecycle. Because maintenance activities can consume 50% to 75% of a project cost, it is important to keep maintenance costs low (vanVliet, 2008). In fact, teams "waste" up to 25% of development time during maintenance, due to technical debt (Martini et al., 2018). Similarly, scientific software projects see maintainability as an important goal because (a) maintenance is costly in terms of productivity and loss of vital scientific work; and (b) Exascale applications are usually written in C, C++, or FORTRAN, which offer high performance but are difficult to evolve and maintain (Schmidberger and Brügge, 2012).
- **Reuse Opportunities**. Productivity is one of the main concerns for scientific software projects (Faulk et al., 2009). One way to improve productivity (i.e., reduce development time) is through software reuse. Because some algorithms are common across projects, especially within a domain, reuse of code should be a helpful approach. As evidence of the potential for reuse, some scientific software projects have explored Software Product Lines, an advanced reuse technique (Costa et al., 2015).

To address these limitations, scientific software developers could benefit from the advances in SE as illustrated in the right side of Figure 1. For example, collaboration among developers and the management of large codebases, could be performed through tools like git and Jenkins (Omar et al., 2014); quality monitoring (especially focusing on Technical Debt) could be performed with SonarQube (Ampatzoglou et al., 2015); and reusability could be facilitated with the use of practices such as refactorings and design patterns (Ampatzoglou et al., 2011). All the above tools could be synchronized by using well-known methodologies for managing software development lifecycles, such as Agile practices (Unhelkar, 2013).

Based on the expected benefits of using SE practices in scientific software development[1], there is a growing interest among scientific software developers to cultivate a culture of SE within their community. This growing interest has begun to impact the literature in this domain. There are papers that report techniques and tools for improving the development of scientific software. Therefore, the goal of this study is to provide a detailed mapping of the current state-of-research and -practice about the use of SE in scientific software development. To properly scope this broad topic, we define three more specific goals: (**G1**) *investigate the SE practices currently used in scientific software development;* (**G2**) *identify the quality attributes that drive the use of SE practices[2]*; and (**G3**) *assess the level of empirical evidence that supports the impact of SE practices on quality attributes*. To achieve these goals, we conducted a Systematic Mapping Study (SMS), focused on classification and categorization of primary studies to provide first understanding of the domain.

## 2. Related Work

This section describes secondary studies (i.e., mapping studies or systematic literature reviews) related to the application of SE practices for scientific software. In Section 2.1 we present an annotated bibliographic reference to such studies, whereas in Section 2.2 we compare them to the current study.

2.1 Detailed Analysis of Primary Studies

---

[1] We expect these benefits to be present for scientific software development in a similar manner as they are for "*traditional*" SE.

[2] In the domain of scientific software development, it might be more realistic to talk about quality expectations

Heaton and Carver's (2015) systematic literature review aimed at identifying claims about how developers of scientific software use SE practices in HPC included 43 papers published prior to May 2015. These papers produced 33 claims about 12 SE practices. They classified each claim based on the type of evidence supporting the claim (e.g., interview or case study). The results suggest that: (a) the most common types of evidence are interviews and surveys, (b) "Issue Tracking and "Version Control" are the SE practices most heavily adopted, and (c) "Verification and Validation" and "Testing" are the practices scientific software developers find important, but are not yet widely adopted. Our current study expands this one by including more recent literature (i.e. beyond May 2015) and expanding the criteria to include SE practices for scientific software development.

Sletholt et al. (2011) literature review about agile practices and their effects on scientific software development investigated (a) the extent to which scientific software projects have used agile practices, and (b) the impact the agile practices have on testing and requirements activities. The review describes 8 papers published between 2000 and 2011. The results suggest scientific projects that adopt agile practices achieve better levels of testing. However, the authors also noticed a positive effect of agile practices on the requirements-related activities. In a follow-up study, which included 5 additional studies, Sletholt et al. (2012) identified 35 agile practices. Of these, 12 originate in Scrum and the rest originate in Extreme Programming. Our current study expands on the work by Sletholt et al. by broadening the scope of the review beyond agile practices in testing and requirements activities.

Farhoodi et al. (2013) performed a systematic mapping study of the most common SE practices for developing scientific software. From the 130 included studies, the main findings were: (a) the majority of scientific software is written in Fortran followed by C++, C, Python, Java and Matlab; (b) the most used SE practices relate to architecture/design, development/coding and testing/verification/validation/quality; and (c) more than one third of the studies do not include validation for the proposed solution. Our current study expands on these results by using more recent papers (this study only includes papers through 2011) and by adding the investigation of quality attributes.

Kanewala and Bieman (2013) presented a systematic literature review to identify the challenges, proposed solutions, and unsolved problems related to testing scientific software. This review includes 62 studies published prior to January 2013. The results include: (a) testing challenges occur due to characteristics of scientific software or to cultural differences between scientific software developers and the larger SE community, and (b) there are techniques scientific software developers can use to overcome some of the testing challenges. Our current paper expands on this work by focusing more broadly than testing and by including more recent papers.

Queiroz and Spitz (2016) performed a systematic literature review to identify a set of UI design practices to support gamification and improve the usability of scientific software. The selection process retrieved 221 primary studies published prior to 2015. The results of this study suggested the "*Lens of the Lab*" as a vehicle to support designers working in collaboration with scientists and software engineers in professional scientific software initiatives. Moreover, the authors proposed that the use of the lens to a project should be a straightforward process, during design stage or consulting appropriate stakeholders about the issues at hand. Our study is broader compared to this of Queiroz and Spitz in the sense that it focuses on SE practices beyond UI design.

Pflüger et al. (2016) conducted a systematic literature review to identify trade-offs between scalability and efficiency on the one hand, and maintainability and portability on the other hand, in simulation software engineering. The selection process retrieved 33 primary studies published between 1990 and 2015. The main findings of this study are: (a) most of the primary studies present some kind of solution or solution proposal; and (b) many of the papers have no clear empirical design, but are opinion pieces or experience reports.

2.2 Comparison to Related Work

Table 1 presents an overview of the papers discussed above, focusing on the research method, the number of included papers, the period covered, and the study goals. The table focus on the three study goals from Section 1 and highlights any goals not included in our study. The research method dictates the depth of analysis in the sense that typically SLRs are more in-depth than SMSs. The number of analyzed papers is an indicator of how broad a study is. The review period highlights how current the results are. The analysis of the goals aims to identify commonalities and differences among the studies. A balance between overlap and novelty is desirable to allow for both comparison and update of results and for novelty to provide additional implications for research and practice.

**Table 1:** Related Work Overview

| Reference | Research Method | # papers | Review Period | G1 - SE Practices | G2- Software Qualities | G3 - Empirical Evidence | Additional Goals |
|---|---|---|---|---|---|---|---|
| Heaton and Carver (2015) | SLR | 43 | through 2015 | X | | X | |
| Sletholt et al. (2011) | SLR | 8 | 2000 - 2011 | X | | | impact of agile to testing and requirements activities |
| Sletholt et al. (2012) | | 5 | | | | | |
| Farhoodi et al. (2013) | SMS | 130 | 1996 - 2011 | X | | X | bibliometrics |
| Kanewala and Bieman (2013) | SLR | 62 | until 2013 | X | | | Definition and challenges of scientific software development |
| Queiroz and Spitz (2016) | SLR | 221 | Until 2015 | partially | partially | | GUI design guidelines Gamification principles |
| Pflüger et al. (2016) | SLR | 33 | 1990-2015 | | partially | X | Trade-offs between the QAs |
| **Our Study** | **SMS** | **359** | **through 2019** | **X** | **X** | **X** | |

Based on Table 1, our study is ***broader*** (contains almost 3 times more papers than the most comparable study that has at least two goals in common) and ***more up-to-date*** (covering almost 5 years from the last review). In terms of goals, our study has the ***widest focus***, since: (a) it does not focus on a specific development methodology like Sletholt et al. (2011), who focus on agile practices; (b) it does not focus on a specific activity like Kanewala and Bieman (2013), who focus on testing; (c) it does not focus on a specific practice like Queiroz and Spitz (2016), who focus on GUI design; and (d) it does not focus on specific quality attributes like Pflüger et al (2016), who focus on four quality attributes. Finally, the main advancement of our work is that it is ***the first study that catalogues the impact of the application of software engineering practices on*** ***quality attributes in scientific software development***; as well as possible trade-offs between qualities.

## 3. Study design

In this section, we present the protocol of the systematic mapping study based on the guidelines described by Petersen et al. (2008).

<u>3.1 Objectives and Research Questions</u>

The goal of this study, expressed in the Goal-Question-Metrics (GQM) format (Basili et al., 1994), is *to analyze the development of scientific software applications for the purpose of characterization and evaluation with respect to the software engineering practices employed and the quality attributes of interest from the point of view of researchers and practitioners*. Based on this goal, we define the following research questions. To address the cross-cutting G3 (*assess the level of empirical evidence that supports the use of SE practices in scientific software development*), we have added sub-research questions in $RQ_2$.

| **$RQ_1$:** Which SE practices used in the development of scientific software have researchers studied the most? |
|---|

The answer to this research question aids scientific software developers in identifying which SE practices researchers have studied most frequently. To further investigate this question, we explore the SE practices used during each development activity (e.g. requirements, design, and testing) and whether there are differences across application domains.

| **$RQ_2$:** Which software quality attributes do researchers study in scientific software development? |
|---|
| $RQ_{2.1}$: Which quality attributes have researchers studied most often for each development activity? |
| $RQ_{2.2}$: What is the impact of SE practices on quality attributes? |
| $RQ_{2.3}$: What is the level of empirical evidence on the aforementioned impact? |

The answer to this research question will expand the knowledge acquired in $RQ_1$, by helping scientific software developers make decisions based both on quality attributes as well as software development activities. The final outcome of this research question will be a 3-fold mapping of practices, activities, and quality attributes. For each of these triplets, the results will produce a value indicating the level of empirical evidence. Scientific software developers can use the outcome of this research question to support their quality planning activities. Researchers can use the results to better scope their future work to address the most important and/or understudied quality attributes.

<u>3.2 Search Process</u>

Based on our goals and research questions, we have chosen to conduct a mapping study rather than a systematic literature review because: (a) the topic is broad, (b) we want to provide a general overview of the topic, (c) the main study goal is developing a classification, and (d) we are not performing a synthesis of results or quality assessment of the primary studies. As searching space, we selected to use the complete content of four well-known Digital Libraries (IEEExplore, ACM, Springer, and ScienceDirect). We chose to search broad Digital Libraries (DLs) rather than specific venues so we could be as inclusive as possible in the selection of papers related to scientific software development and SE. Figure 2 provides an overview of the process, which is organized into four steps: (a) searching Scopus without a start date; (b) filtering results to retain only the studies published in the 4 DLs; (c) removing duplicates; and (d) applying the inclusion / exclusion criteria. In the end, we retained 359 primary studies to include in this mapping study.

**Figure 2:** Overview of Search Process

In more detail: first we developed a search string (see box below) to identify papers relevant to SE AND scientific software development. Because scientific software often demands a large number of calculations over vast amounts of data, these applications make heavy use of High-Performance Computing (HPC). In fact, more than 70% of HPC applications address problems outside of computer science (Schmidberger and Brügge, 2012). Therefore, to be as inclusive as possible, we included "HPC" in the list of search terms. We performed this search on the title / abstract / keywords of all papers in Scopus, which includes papers from the four DLs of interest. We used Scopus rather than the DL search engines to avoid inconsistency issues and problems identified in other studies (e.g., Springer allows searching of only one field: full text or title—but not abstract).

> ("software engineering" OR "software development" OR "software practice")
> AND
> ("scientific computing" OR "scientific software" OR "computational software" OR "scientific programming" OR "high performance computing" OR "high performance science" OR "HPC" OR "research software engineering" OR "research software development")

Second, we manually filtered the results to retain studies published in the four DLs. Third, we removed duplicated papers. Finally, since we used "HPC" in the search string, we had to ensure that we include only papers relevant to scientific software development. Thus, we defined the following Inclusion Criteria:

- IC1: The primary study is applied in scientific domain;
  AND
- IC2: The primary study defines/uses one or more SE practices; OR
- IC3: The primary study evaluates one or more SE practices; OR

- IC4: The primary study uses one or more quality attributes; OR

In other words, the final inclusion criterion is: IC1 AND (IC2 OR IC2 OR IC4). Similarly, we defined the following exclusion criteria. We excluded a study if it met at least one of them.

- EC1: The primary study is written in a language other than English.
- EC2: The primary study is an editorial, keynote, biography, opinion, tutorial, workshop summary report, progress report, poster, or panel.
- EC3: The primary study is of horizontal perspective covering the complete spectrum of SE practices (these studies have been reported as related work).
- EC4: The paper is focusing on HPC, without a reference to scientific software development

During the inclusion/exclusion process, two authors independently examined 921 studies. In particular, the first two authors inspected the publications' full text and assigned a score on a 4-point scale (4: strong inclusion, 1: strong exclusion)—leading to a maximum score of 8 points. Following the threshold used by Farhoodi et al. (2013), we retained studies that had a score of 6, 7 or 8. For the studies that scored 5 (88 cases), the third author reviewed them and made a final decision.

3.3 Data Extraction

During the data extraction phase, we collected a set of variables that describe each primary study. The complete dataset is available online[3]. To strengthen the validity of data extraction, we used the following systematic process. The first two authors independently extracted data. If there were inconsistencies in the extracted information, the authors first discussed the inconsistencies. If they were not able to resolve the discrepancies, the third author joined the discussion to resolve the disagreement. For every study, we extracted and assigned values to the following variables:

- [V1] **Title**: *title* of the paper.
- [V2] **Author**: *list of authors* of the paper.
- [V3] **Year**: *publication year* of the paper.
- [V4] **Type of Paper**: whether the paper appears in a *conference* or *journal or workshop*.
- [V5] **Publication Venue**: *name* of the corresponding journal or conference.
- [V6] **Development Activity**: development activity investigated in the primary study (e.g., *requirements*, *architecture*, *design*, *implementation*, *testing*)
- [V7] **Type of Software Artifact**: software artifacts mentioned in the study (e.g. class diagram, use case, etc.)
- [V8] **Names of Software Engineering Practice**: SE practices described in the paper (e.g. design patterns, traceability, model-driven development etc.)
- [V9] **Names of Quality Attributes**: *list of the names of quality attributes* investigated in the study, exactly as reported in the primary study.
- [V10] **Programming Language**: programming language used in study (e.g. *Fortran, C, C++ etc.*)
- [V11] **Application Domain**: application domain in which the software is used (e.g. astronomy, geology, chemistry, etc.)

---

[3] https://se.uom.gr/wp-content/uploads/SLR_HPC_SE.xlsx

[V12] **Empirical Research Method**: the type of empirical method (e.g., case study, survey, experiment, action research, ethnography, field research) used to validate the impact of the SE practice on the Quality Attribute.

[V13] **Impact:** The outcome of the empirical validation (positive, negative, or neutral). In cases when the paper studied more than one QA and the impact on one QA was positive and the impact on the other QA was negative, we marked the study as a ***trade-off*** (Feitosa et al., 2015). In cases when a blended (either positive or negative) impact was identified, based on some parameter, we marked the study as a ***cut-off*** (Charalampidou et al., 2017).

For all variables, we performed data extraction based on the terminology used in the primary study. In other words, we did not try to change the terms if we believed the authors used an incorrect term. We did not have a pre-determined list of development activities. But rater allowed the reported activities to emerge from the data. The identification of development activities and their mapping to software artifacts is not a trivial task, due to the existence of various processes and Software Development Lifecycle Models (SDLC). To catalogue activities and perform the mapping of artifacts to activities, we used a number of sources, as there was no single source that contained all datapoints (artifacts or activities) that we have identified. Specifically, we used four process models (RUP, OpenUP, ICONIX, and Scrum) and the IEEE 830 Standard. For example, the "*Software Architecture Document*" is mapped to the architecture activity according to RUP (RUP calls activities as workflows), whereas the term "*Use-Case Model*" is mapped to the requirements activity, based on both OpenUP and ICONIX (OpenUP calls activities as domains while ICONIX calls them disciplines. In cases of artifacts that can be mapped to more than one activity depending on the SDLC model, we map the artifact to the activity that produces it. For example, "*reported bugs/issues*" can be treated as parts of testing (in RUP) or requirements (since they are fed as backlog items in SCRUM). We map them to testing, because initially, bugs are considered as an outcome of testing and later are fed back to the system as requirements.

**Table 2:** Development of Empirical Methods Classification Schema

| Research Methods | Easterbrook et al. (2008) | de Magalhaes (2014) | Hummel (2014) | Silva et al. (2015) | Stol et al. (2009) | ESEM Conference | Count |
|---|---|---|---|---|---|---|---|
| Survey | X | X | X | X | X | X | 6 |
| Case Study | X | X | X | X | X | X | 6 |
| Action Research | X | X | X | X | X | X | 6 |
| Experiment | X | X | X |   | X | X | 5 |
| Ethnography | X | X | X |   | X |   | 4 |
| Field Research/Study |   | X |   |   | X | X | 3 |
| Grounded Theory |   |   | X |   | X |   | 2 |
| Simulation |   |   | X |   |   | X | 2 |
| Quantitative Analysis |   |   |   |   | X | X | 2 |
| Experience Report |   |   |   | X |   | X | 2 |
| SLR |   |   | X |   |   | X | 2 |
| Theoretical/Descriptive |   |   |   | X |   |   | 1 |
| Meta-Analysis |   |   |   |   |   | X | 1 |
| Qualitative |   |   |   |   |   | X | 1 |

| Research Methods | Easterbrook et al. (2008) | de Magalhaes (2014) | Hummel (2014) | Silva et al. (2015) | Stol et al. (2009) | ESEM Conference | Count |
|---|---|---|---|---|---|---|---|
| Focus Group | | | X | | | | 1 |

Conversely, for [V12], we reused a list of empirical methods (Charalampidou et al., 2020). In particular, we considered several names of empirical research methods found in literature, as shown in Table 2. The first column of the table shows the research method names, while the next 6 columns indicate the sources that consider the method as empirical research. To identify the list of sources, we began with the most well-known papers and books dealing on empirical software engineering research (e.g., (Wohlin et al., 2012) and (Runeson et al., 2012)). However, these sources focused on specific research methods (i.e., experiments and case studies respectively). Thus, we identified five papers that focused on empirical research from a more generic perspective. Additionally, since we collected the empirical research methods listed in the call for papers from the *International Symposium on Empirical Software Engineering and Measurement (ESEM)* is the main conference for the empirical SE community. Similarly, we examined the aims and scope of the journal *Empirical Software Engineering*, however we did not identify additional keywords. The last column of the table shows how many times each research method appears as an empirical approach in the six sources. In our classification schema, we only retained methods that appeared in at least two sources (green cells). Note that although the term SLR had two references, we did not include it in our framework because it is not a primary study.

In reviewing the primary studies, we identified the empirical method as follows: (a) in cases where the study explicitly mentioned the study type, we validated that the empirical setup matched the term and then assigned it to the corresponding variable; and (b) in cases where the study did not explicitly mention the empirical method, we determined it based upon the study design.

3.4 Data Analysis

We collected variables [V1] – [V5] for documentation reasons. We use variables [V10] and [V11] for demographics. Table 3 provides a mapping between the research questions and the remaining variables, along with the type of analysis performed on the data. For $RQ_1$ and $RQ_2$, we provide the frequency table of variables [V8] and [V9], respectively.

**Table 3:** Mapping of paper attributes to RQs

| Research Question | Variables Used | Analysis Method |
|---|---|---|
| $RQ_1$ | [V6], [V7], [V8] | Crosstabs for [V6], [V8], Crosstabs for [V7], [V8] |
| $RQ_{2.1}$ | [V6], [V9] | Crosstabs for [V6], [V9] |
| $RQ_{2.2}$ | [V8], [V9], [V12] | Crosstabs for [V8], [V9], Crosstabs for [V8], [V12] |
| $RQ_{2.3}$ | [V8], [V9], [V12] | Crosstabs for [V8], [V12], Crosstabs for [V9], [V12] |

Due to the large number of SE practices in the literature we performed pre-processing. To consolidate and merge similar values of [V8] we used Open Card Sorting (Spencer 2009). In particular, we: (a) identified more generic practices (i.e., super-categories) from the SE practices in the primary studies—e.g., we developed a theme "*Programming Technique*"; (b) reviewed the themes to find candidates for merging—e.g., we mapped "*Model-Driven Engineering*" as "*Programming Technique*"; and (c) defined the names of the final themes and super-categories. In the manu-script we report on super-categories, but in the dataset, we report the more detailed categories. The first author performed the process. Then the second and third authors validated the results.

**4. Results**

In this section we present the results and some initial interpretation of our data. In particular, Section 4.1 presents an overview of the studies based upon frequencies of demographic information. Then Sections 4.2 and 4.3 provide the answers to $RQ_1$ and $RQ_2$, respectively.

4.1 Overview of Included Studies

As a general overview, we provide the frequencies of the main variables describing the primary studies. We note that from all presented frequency tables, we have excluded items with one occurrence. Additionally, datapoints, which split with slash, correspond to merged datapoints, whereas datapoints, which are split with comma, correspond to different ones with the same count. First, Figure 3 illustrates the number of studies published per year. Based on these numbers it appears that in '80s and '90s, scientific software development research was not particularly focused on the use of SE practices. However, after 2000 (and in particular after 2004), the number of studies has increased substantially. This result is not surprising given that scientific software has increased in size, complexity, or the need for other properties for which SE practices can be helpful (e.g. scalability and portability).

**Figure 3:** Frequency of Publication

Table 4 lists the frequencies of studies that investigated specific development activities. Table 4 shows that the most frequently studied development activity is *implementation*, followed by *architecture* and *testing*. Interestingly, only less than 2% of the studies investigated the *maintenance* activity, despite the fact that maintenance is an important and costly activity for software that evolves over time. The prevalence of *implementation*, *testing*, and *architecture* activities in scientific software development is consistent with findings from earlier studies (Farhoodi et al. 2013), (Odun-Avo et al. 2018), and (Heaton and Carver 2015).

**Table 4:** Frequency of Development Activities

| Development Activity | Count | Development Activity | Count |
| --- | --- | --- | --- |
| Implementation | 159 | Project Management | 14 |
| Architecture | 71 | Maintenance | 7 |
| Testing | 50 | Deployment | 6 |
| Design | 28 | Integration | 4 |
| Requirements | 23 | | |

The results in Table 5, which lists the software artifacts studied in the primary studies, show that *source code* is the most frequently studied software artifact (45%). This outcome is expected, because it follows the distribution in Table 4: i.e., artifacts that are produced by popular activities in Table 4 score higher in Table 5, as well (with the same ranking for the first five artifacts).

**Table 5:** Frequency of Software Artifacts

| Software Artifact | Count | Software Artifact | Count |
|---|---|---|---|
| Source Code | 162 | Domain Model / Class Diagram | 19 |
| Component / Component Diagram | 61 | Requirements | 14 |
| Unit Test / Test Case / Test Plan | 32 | Flow Charts, Use Case Diagrams | 9 |

Table 6 provides a cross-tabulation of the results from Tables 4 and 5. These results show that *source code* appears in all development activities[4]. In addition, for most of the activities, the artifact that is most associated with the activity, based on SDLC models, is the most prevalent (e.g. the *requirements* artifact in the *requirements* activity and *component diagram* in the *Architecture* activity).

**Table 6:** Frequency of Software Artifacts per Development Activity

| Development Activity | Software Artifact | Freq. |
|---|---|---|
| Implementation | Source-Code | 155 |
| Architecture | Component/Component Diagram | 61 |
| Testing | Unit Test/Test Case/Test Plan | 32 |
| | Source-Code | 8 |
| Design | Domain Model/Class Diagram | 17 |
| | Use Case Diagram, Flow Chart | 4 |
| Requirements | Requirement | 14 |
| | Domain Model/Class Diagram | 2 |
| Maintenance | Source-Code | 3 |

Table 7 lists the programming languages reported in the included studies. The finding that C++, C, and Fortran are the most common languages for scientific software development is consistent with prior studies ((Johanson et al., 2018), (Farhoodi et al., 2013), and (Amaral et al., 2019)). This result is expected because Fortran is still a dominant language for large-scale scientific applications that heavily rely on mathematical operations (Faulk et al. 2009). The fact that the

---

[4] To see details about architecture, design, and requirements see Appendix B.

C-family languages are ranked first, can be explained by the general popularity of the languages[5] and the curricula of most natural science departments, which first acquaint developers of scientific software with C and C++.

**Table 7:** Frequency of Programming Languages

| Programming Language | Count | Programming Language | Count |
|---|---|---|---|
| C++ | 81 | Python | 22 |
| C | 51 | Matlab | 14 |
| Fortran | 44 | Corba, R | 4 |
| MPI | 31 | Mathematica, Cuda, Cell, OpenCL, JS, Ruby | 2 |
| Java | 24 | | |

As reported in Table 8, researchers involved in scientific software development come from a variety of domains. Note that the sum of the studies in Table 8 is less than the total included studies because several studies did not report application domain. The most frequent domains reported in Table 8 are those that are in need of large-scale simulations, which often require large amounts of computational power and process large amounts of data. The results of the study are in agreement with previous work (Farhoodi et al., 2013) who suggested that physics, biology and mathematics are the top application domains of scientific software development.

**Table 8:** Frequency of Application Domain

| Application Domain | Count | Application Domain | Count |
|---|---|---|---|
| Biology | 12 | Airborne | 6 |
| Physics | 12 | Chemistry | 5 |
| Mathematics | 11 | Medical | 3 |
| Climate / Environment | 9 | Communications, Green Computing | 2 |
| Geoscience / Cosmology | 8 | Music, Material Science | 1 |

Table 9 lists the empirical methods authors used to validate the proposed approaches. An interesting finding is that a large percentage of studies (~33%) have no validation of the proposed methods. For the studies that did use an empirical validation, the distribution mirrors traditional SE, in which case studies and experiments are the dominant type of research (Molleri et al., 2019). An interesting observation from the data is that prior to 2000, very few studies applied empirical methods (less than 20%). However, after 2000, researchers begun to more frequently validate their results via empirical methods (more than 50%).

**Table 9:** Frequency of Empirical Methods Used for Validation Purposes

| Empirical Method | Count |
|---|---|

---

[5] https://www.tiobe.com/tiobe-index/

| | |
|---|---|
| Case Study | 76 |
| No Validation[6] | 57 |
| Experiment | 32 |
| Survey | 9 |
| Ethnography | 1 |

4.2 RQ$_1$ –Software Engineering Practices in Scientific Software Development

In this section, we present the results of our mapping study related to SE practices in scientific software development. Table 10 lists the software practices most frequently reported in the included studies (after the consolidation process described in Section 3.6). The results are consistent with what one would expect given the nature of scientific software development.

**Table 10:** Frequency of Software Practices (RQ$_1$)

| Software Practice | Freq. | Software Practice | Freq. |
|---|---|---|---|
| Reuse or Libraries or API | 41 | Quality Assurance, with or without Metrics | 17 |
| Programming Technique | 33 | Agile Practices | 15 |
| Parallel Programming or Distributed Software | 25 | Quality Optimizations | 15 |
| Software Development Process Improvement or Lifecycle Management | 24 | (Introduce or Use a Specific) Programming Language | 14 |
| Component-Based Software Development | 21 | Integrated Development Environment (IDE), Domain Specific Languages | 13 |
| Development Framework (propose) | 21 | Code Generation | 13 |
| Testing, Regression or Automated Testing, Testing without oracle | 19 | Project Management, Formal Testing Methods, Software Architecture, Requirements Specification | 11 |
| Design and Architecture Models | 18 | People Management or Communication | 9 |

The results suggest that the most commonly reported practices are related to implementation. The most common SE practice is software reuse, not in the form of source code reuse, but in the form of developing an artifact for reuse (Lambropoulos et al., 2018). The most common packaging for this type of software is a library that can solve common problems in a domain. In some cases, researchers discuss how Application Programming Interfaces (APIs) can ease reuse of these third-party libraries (Zaimi et al., 2015). Second, we observed that a study discusses a variety of **Programming Techniques**, including: "*Model-Driven Engineering*", "*Skeleton Programming*", "*Task Scheduling*", or programming paradigms (e.g., "*Aspect-Oriented Programming*"—AOP or "*Object-Oriented Programming*"—OOP). The discussion of programming techniques (none of which appeared in more than 3% of included papers), suggests that the scientific software development community has high interest on how to achieve programming efficiency. Third, we observe some architecting practice, such as **Parallel or Distributed Software** Architectures and the dominant practice of higher-level reuse, i.e., **Component-Based Software Engineering**

---

[6] We note that from the table we excluded the studies (51%) that do not refer to a specific QA, since the empirical method variable cannot be defined.

***(CBSE)***. The focus on parallelization can be attributed to the need for execution performance of the very complex calculations usually performed in scientific software. The focus on CBSE suggests an attempt to systematize reuse in earlier phases of development. Fourth, a large portion of research is spent on ***Testing***. Because scientific software applications are complex, developers must perform different types of testing to verify and validate the results. Furthermore, there were seven additional papers that discussed *test-driven development* and *quality assurance*, which could both be considered an aspect of testing. Finally, we need to underline the interest on practices related to ***Software Process Improvement***. As scientific software developers learn more about SE practices, it makes sense that studies about how to best assemble those practices into a lifecycle could benefit software development. For instance, we have observed an interest in *"Collaborative Development"*. The prevalence of this topic is consistent with the fact that the complexity of most scientific projects requires collaboration among multiple developers, often with diverse backgrounds.

***SE Practices per Activity:*** In Figure 4 we present a map between activities and SE practices.

(a) Implementation

(b) Architecture

(c) Testing

(d) Design

(e) Requirements          (f) Project Management

**Figure 4:** Frequency of Software Practices per Development Activity ($RQ_{1.1}$)

From the results we can make the following observations:
- During implementation, developers of scientific software are interested in the development and reuse of code through libraries, the adoption of programming techniques, the use of development frameworks, the collaboration between the developers, and in different techniques and methods for improving the quality of the software.
- During the architecture/design activity, developers of scientific software are interested in reuse through component-based software engineering, the use of models and techniques for improving the quality of the software, and reducing the cost of the software (e.g., SPL).
- Researchers have studied the use of multiple types of testing in software development.
- During the requirements activity, developers of scientific software focus on gathering requirements using various techniques (e.g., interviews, workshops, etc.) from stakeholders and building a better understanding for the requirements.
- During the project management activity, it seems that developers of scientific software are focused on peopleware aspects, e.g., practices for managing the human factors to deliver projects consistently, efficiently, and on time and within budget.

We note that the sum of the items in Figure 4 may be larger than reported in Table 10 because some studies linked an SE practice to more than one activity. An interesting result is that, while a previous study reported testing as one of the most understudied activities in scientific software development (Heaton and Carver, 2015), we found, five years later, testing is now one of the highest studied activities.

Finally, we note that some SE practices are cross-cutting, in the sense that they can be applied in more than one development activities. For example, "*Quality Assurance with or without metrics*" can be performed during implementation quality assurance through code reviews or linters. At the architecture / design phase it can be performed with design reviews or inspections. Finally, at testing phase it can be performed through coverage metrics. However, the number of cross-cutting SE practices is not substantial enough to perform an analysis for checking differences among different development activities.

***SE Practices per Application Domain***: Table 11 presents the cross-tabulation of SE practices and application domains. Based on the findings of Table 11, we have observed that there are no differences in the SE practices that are applied across different application domains. Thus, for the rest of this manuscript, we report on the dataset as a whole, without differentiating between application domains, or development activities.

**Table 11:** Frequency of Software Practices per Application Domain

| Domain | Software Practice | Freq. | Domain | Software Practice | Freq. |
|---|---|---|---|---|---|
| Biology | Reuse or Libraries or API | 2 | Climate / Environment | DSL | 2 |
| | People Management or Communication | 2 | | Release as OSS | 2 |
| | Agile Practices, Software Architecture, GUI Design, Programming Language, Development of Compilers, Requirements Specification, Parallel Programming or Distributed Software, Requirements Elicitation, Design and Architecture Modelling | 1 | | IDE, People Management or Communication, Software Development Process Improvement or Lifecycle Management, requirements elicitation, Reuse or Libraries or API, Separation of Concerns | 1 |
| Mathematics | Programming Technique | 3 | Chemistry | Reuse or Libraries or API | 2 |
| | Code Generation | 3 | | Agile Practices, Programming Language, Requirements Elicitation | 1 |
| | Separation of Concerns, Requirements Management, Documentation, Software Development Process Improvement or Lifecycle Management, People Management or Communication, Development Framework (Propose), IDE | 1 | Airborne | Quality Assurance, with or without Metrics, Release as OSS, Programming Language, Design and Architecture Modelling, Design and Architecture Models, Requirements Specification, Development Framework (Usage), Development Framework (Propose), Cost / Effort Estimation | 1 |
| Physics | Development Framework (Propose) | 3 | Geoscience / Cosmology | Reuse or Libraries or API | 2 |
| | Project Management | 2 | | Quality Optimizations | 2 |
| | People Management or Communication | 2 | | Requirements Elicitation | 2 |
| | Reuse or Libraries or API, Quality Optimizations, Risk Management, Agile Practices, Component-Based Software Development, Requirements Elicitation, Parallel Programming or Distributed Software, Design and | 1 | | Collaborative Software Development/Version Control/Configuration Management, Programming Technique, Requirements Specification, Release as OSS, Quality Assurance, with or without Metrics | 1 |

| Domain | Software Practice | Freq. | Domain | Software Practice | Freq. |
|---|---|---|---|---|---|
| | Architecture Modelling | | | | |

4.3 RQ₂ –Quality Attributes relevant to Scientific Software Development

In response to RQ₂, Table 12 lists the frequencies of quality attributes targeted by the primary studies. **Performance** is the most studied quality attribute. This result makes sense because developers of scientific software need to obtain their data and/or analysis results as quickly as possible (García et al., 2013). The second ranked quality attribute is **Productivity**, which refers to development efficiency: explaining the focus on reuse (libraries and CBSE)—interestingly, reusability is ranked low. This indicates that the goal of scientific software developers is not on systematic reuse, but on reducing development time. The following attribute is **Maintainability**, which is a rationale outcome in the sense that such applications change frequently, and therefore it is desirable to reduce the effort for updating the software. Finally, an interesting observation is that portability is the fourth most frequently studied quality attribute, since developers of scientific software are interested in developing applications that are portable to parallel development environments (Watson and De Bardeleben, 2006).

**Table 12:** Frequency of Targeted Quality Attributes

| Quality Attribute | Freq. | Quality Attribute | Freq. |
|---|---|---|---|
| Performance | 67 | Robustness (also referenced as Fault Tolerance) | 12 |
| Productivity | 33 | Complexity (also referenced as Understandability) | 10 |
| Maintainability (also referenced as Extensibility or Flexibility or Changeability) | 28 | Interoperability, Usability | 7 |
| Portability | 26 | Energy and Memory Efficiency | 6 |
| Scalability | 24 | Security or Safety | 5 |
| Correctness (also referenced as Accuracy or Reliability) | 18 | Modularity, Testability (also referenced as Verifiability) | 3 |
| Reusability | 13 | | |

In response to RQ$_{2.1}$, Figure 5 presents the QAs of interest for each development activity.

(a) Implementation                    (b) Architecture

(c) Testing
(d) Design

(e) Project Management
(f) Deployment

**Figure 5:** Frequency of Quality Attributes per Development Activity (RQ$_{1.1}$)

Based on this information, we can note that *Maintainability* is of interest in all development activities except for deployment. As explained before, scientific software developers are interested in efficiently maintaining their code because of the need to make minor adjustments across versions and perform corrective maintenance. These needs have made maintainability important across multiple development activities, including early activities like project management and architecture, when the impact of considering maintenance can be far-reaching.

In response to RQ$_{2.2}$, Table 13 lists the quality attributes associated with each of the top-20 most common software practices. In a parenthesis (when applicable) we denote the number of studies that report a negative or neutral impact of the SE practice on the quality attribute. We note that as negative we also designate cases in which a study reveals that the effect is not uniform, i.e., there are cases when the SE practice has a positive effect and others that it is negative. Based on the results, we can claim that researchers publish positive results more frequently than negative results (less than 2% of the studies report negative results). This finding is expected due to the phenomenon termed *publication bias* (Ampatzoglou et al., 2019). The tendency to publish only positive results has also been identified in traditional software engineering research. Thus, the number of venues that explicitly state (in their call for papers) that they accept negative results is increasing.

Additionally, in traditional software engineering, there is no design decision or application of a practice that does not come without a cost (Ampatzoglou et al., 2021). Thus, any decision-maker needs to consider various quality attributes and explore possible quality trade-offs between them (i.e., one QA is improved, whereas another deteriorates) (Bass et al., 2003). By seeking for explicit trade-off analysis studies in our dataset, we have identified only one (Naughton et al., 2018) study that identifies trade-offs and only two ((Abdullin et al., 2017) and (Sapuan et al., 2018)) that identify cut-off points (i.e., the same practice can have both positive and negative impact, based on some parameters).

Table 13: Frequency of Quality Attributes per Software Practice

| Software Practice | Quality Attribute | Freq. |
|---|---|---|
| Reuse or Libraries or API | Performance | 9 |
| | Portability | 3 |
| | Scalability | 3 |
| | Robustness (also referenced as Fault Tolerance), Productivity, Reusability, Maintainability (also referenced as Extensibility or Flexibility or Changeability) | 2 |
| Programming Technique | Performance | 7 |
| | Maintainability (also referenced as Extensibility or Flexibility or Changeability), Productivity | 3 |
| | Scalability | 3 |
| | Reusability, Complexity (also referenced as Understandability), Correctness (also referenced as Accuracy or Reliability) | 2 |
| Parallel Programming or Distributed Software | Performance | 8 (1) |
| | Portability | 5 |
| | Scalability | 4 |
| | Correctness (also referenced as Accuracy or Reliability), Maintainability (also referenced as Extensibility or Flexibility or Changeability) | 3 |
| | Productivity | 2 |
| Software Development Process Improvement or Lifecycle Management | Performance | 5 |
| | Productivity, Maintainability (also referenced as Extensibility or Flexibility or Changeability), Portability | 3 |
| | Usability | 2 |
| Component-Based Software Development | Performance | 8 |
| | Portability | 3 (1) |
| | Maintainability (also referenced as Extensibility or Flexibility or Changeability), Reusability | 3 |
| | Modularity, Scalability | 2 |
| Testing, Regression or Automated Testing, Testing without oracle | Performance, Correctness (also referenced as Accuracy or Reliability), Portability, Scalability, Productivity | 2 |
| Development Framework (propose) | Portability, Performance | 4 |
| | Productivity, Scalability | 3 |
| | Maintainability (also referenced as Extensibility or Flexibility or Changeability), Energy or Memory Efficiency, Correctness (also referenced as Accuracy or Reliability), Interoperability, Robustness (also referenced as Fault Tolerance) | 2 |
| Quality Assurance, with or without Metrics | Productivity, Complexity (also referenced as Understandability) | 2 |
| Design and Architecture Models | Performance | 3 |
| | Reusability, Maintainability (also referenced as Extensibility or Flexibility or | 2 |

| Software Practice | Quality Attribute | Freq. |
|---|---|---|
| | Changeability) | |
| Programming Language | Performance | 6 (2) |
| | Productivity | 4 |
| Quality Optimizations | Maintainability (also referenced as Extensibility or Flexibility or Changeability) | 5 |
| | Performance | 4 |
| | Productivity | 2 |
| Agile Practices | Maintainability (also referenced as Extensibility or Flexibility or Changeability), Performance, Security, Safety | 2 |
| | Productivity | 2 (1) |
| Domain Specific Language | Reusability | 3 |
| | Productivity | 2 |
| Code Generation | Performance | 4 (1) |
| | Productivity | 2 |
| Project Management | Productivity | 3 |
| Formal Testing Methods | Robustness or Fault Tolerance | 2 |
| | Performance | 2 (1) |
| Software Architecture | Performance | 4 |
| People Management or Communication | Productivity, Robustness (also referenced as Fault Tolerance) | 2 |

In response to RQ$_{2.3}$, we present two views of empirical validation methods. In Figure 6, we present a bubble chart representing the frequency with which each QA is evaluated by every empirical method. In Figure 7, we visualize the frequency with which the impact of each SE practice has been validated.

Figure 6 provides an overview of the level of validation that exists for each pair. The results show that: (a) researchers have studied performance (18%), scalability (28%), correctness (29%), and complexity (10%) most rigorously, based on the lowest percentage studies without any validation (excluding the quality attributes with only one study); and (b) reusability (54%), usability (50%), energy consumption (50%), and robustness (42%) need more empirical evidence because they have the lowest percentage of validation studies.

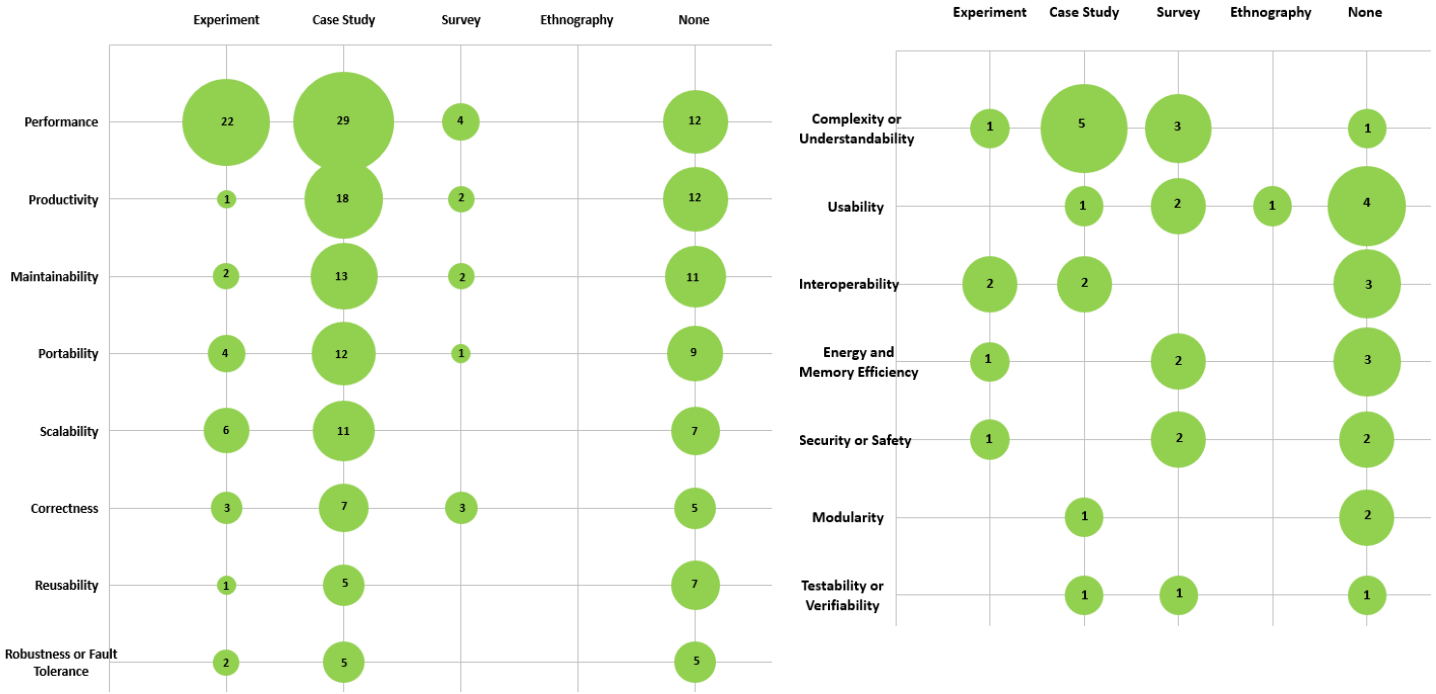

**Figure 6:** Frequency of Empirical Methods for the Validation of QAs

Based on Figure 7, we can draw some observations. By focusing on the extreme cases: two out of the three studies that address "*People Management*", 42% of studies that focus on "*Software Lifecycle Improvement*" (SDLC) and 40% of studies that study "*Software Architectures*" do not provide any empirical validation. One possible explanation for this observation is that earlier stage activities (in which not many artifacts have been developed) may be more difficult to evaluate; due to the need of more advance qualitative approaches. On the other hand, 94% of studies that focus on "*Parallelization or Distribution*" of code have rigorous validation. This result can be explained by the fact that in most of the cases, the performance indicator for this practice is the time required to execute the software, which is relatively easy to obtain.

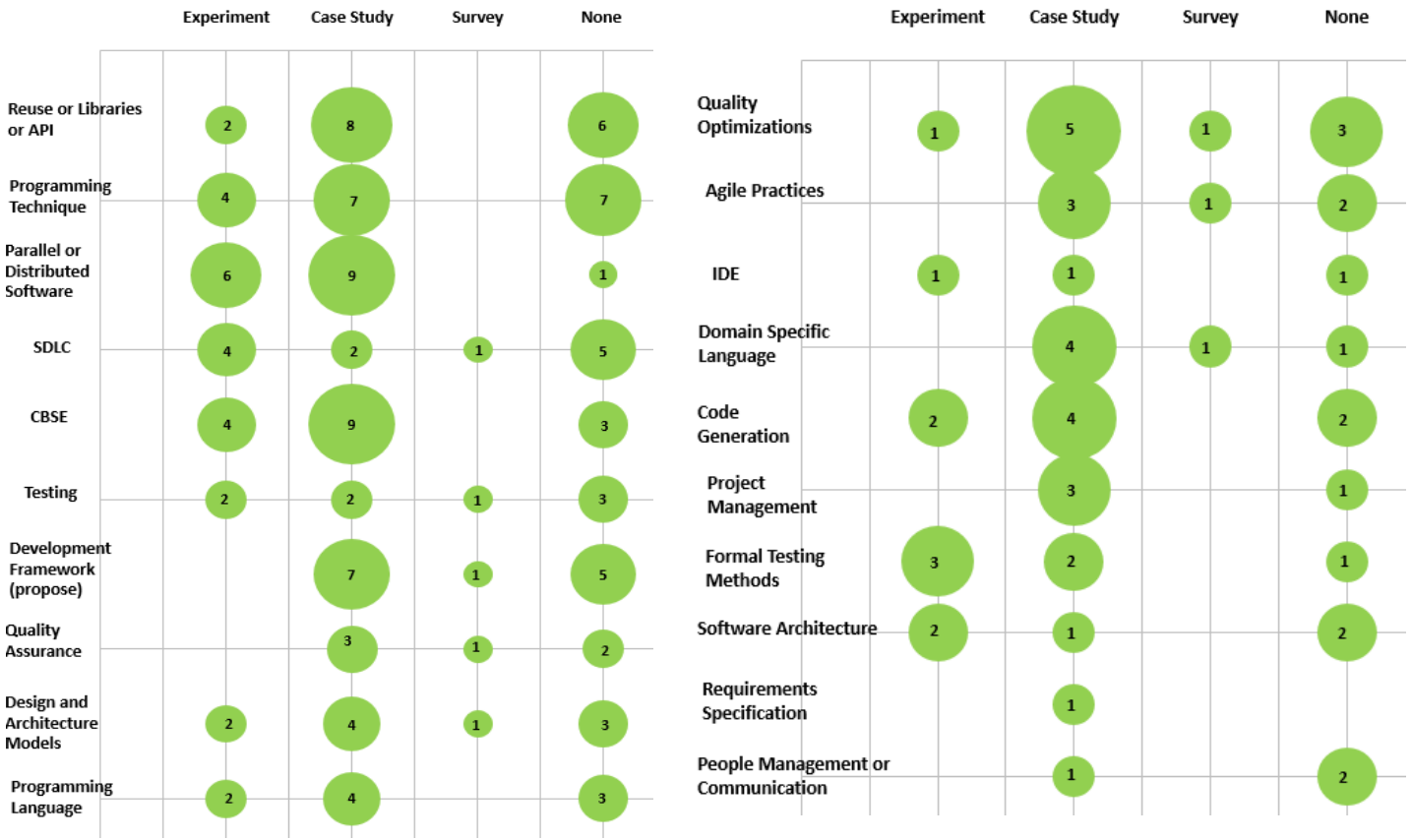

**Figure 7:** Frequency of Empirical Methods for the Validation of SE Practices

## 5. Discussion

### 5.1 Overview and Interpretation of Results

The results of our mapping study are consistent with prior studies. The software practices identified as the most studied match prior results as follows: Odun-Avo et al (2018) also highlighted the importance of *process improvement*, multiple studies noted the importance of *testing* (Heaton and Carver, 2015), (Johanson et al., 2018), and (Farhoodi et al., 2013), as well as multiple studies noted the importance of focusing on *application development* and *programming languages* (Odun-Avo, et al., 2018), (Assiroj, et al., 2018), and (Farhoodi, et al., 2013). By having a closer look at the more frequently used SE practices, one can observe that their majority deals with implementation and testing tasks:

- ***Ease of Development***. Papers report multiple practices that aim to ease and speedup software development including *reusable code identification algorithms*, *use of 3rd party libraries*, *software development frameworks*, and *developing applications that can evolve*. This emphasis can be explained by the fact that often scientific applications share common functionalities, therefore, reuse is an appealing way of improving productivity. A similar productivity increase can be achieved through the use of tools, frameworks and libraries.
- ***Testing***. Papers reported two practices related to testing: *test-driven development* and *specific types of testing* (e.g., regression and automated testing). This outcome suggests that scientific software developers see a need to produce correct results, especially in cases of simulations related to critical software systems. For example, when implementing optimization algorithms (e.g., simulation of $CO_2$ emission (Damartzis et al., 2018)), scientific software developers prioritize finding a solution that is a global optimum rather than a local one over the time required to produce that solution. To ensure this goal, the developers need exhaustive validation.

- **Coding**. In addition to practices that aim at making development easier, more efficient, and more productive, another research direction is the study of *programming languages*, *compilers*, *code generation*, and *code management strategies*. All the identified practices suggest that scientific software developers view coding tasks as very important, yet still struggle with them. Therefore, researches invest substantial effort in providing scientific software developers with tools that support coding.
- **Project Management**. The scientific software domain seems to be in need of project management practices. To this end, *software development processes* are the 2$^{nd}$ most studied SE practice. Such measures that do not focus on a specific activity (e.g., the application of standardized development practice, rather than an ad-hoc one) is an interesting topic for scientific software researchers, because they expect these practices to yield various benefits (e.g., higher quality software and increased productivity).
- **Quality Assurance**. Finally, the literature shows that scientific software researchers are interested in software quality assurance procedures that focus on non-functional requirements, in addition to testing (described above), which focus on functional requirements.

By further focusing on the most-studied in the scientific software development literature, we see that *performance* is the top-priority for scientific software developers. The next quality attributes of interest appear to be *maintainability* and *development productivity*. This finding suggests that developers of scientific software are highly interested in decreasing the effort they spend in software development. In particular, they need solutions that make software construction more productive, but also decrease their maintenance effort, since they appear to deem it as non-negligible. Finally, the next quality attributes are those that are primarily improved by the used practices (*reusability*, *correctness*, and *reliability*), the *size* of the software (probably as a proxy of effort as well), and *portability*. Portability is a sub-characteristic of maintainability that seems important in scientific software development, in the sense that it is related to performance. For instance, to decrease execution time in many cases solution architects move certain calculations to GPU (from CPU). To ease such a tailoring the original code must be portable to other types of processors, hardware, operation systems, etc. The results on the importance of quality attributes comply with those of Johanson et al. (2018), who suggested that the main quality drivers of scientific software applications are performance, maintainability, portability, and correctness.

5.2 Implications to Researchers and Practitioners

The main benefit of this study for practitioners is guidance in selecting the most fitting SE practices for improving quality attributes for each development activity. To illustrate how a practitioner can use the results, we first need to develop a synthesized view of the results.

- *Selection of development activity*. First, the practitioner needs to specify the software development activity of interest. Based on this selection he/she can identify the most studied SE practices for this activity from the answer to $RQ_{1.1}$ in Figure 4.

- *Selection of quality attributes*. Then, to limit the SE practices of interest, the developer needs to identify which practices are related to the quality attribute of interest, for each development activity based on the answers to RQ$_{2.1}$ and RQ$_{2.2}$.

Figure 8 presents a portion of a matrix we developed to illustrate the results. It includes development activities, SE practices, and quality attributes (as dimensions) and inside each cell the effect of the practice on the quality attribute. Due to the size of the matrix, we show only an example here. The full matrix is available in Appendix C. To reduce the complexity of the representation, we mapped development activity and SE practices in the matrix rows via nesting. The first two columns present the most studied SE practices for each development activity. Then, in the remaining columns, we highlight which SE practices are related to each quality attribute. Developers can use this figure to choose which SE practices are most relevant when developing scientific software.

| Activity | Software Practice | Performance | Productivity | Maintainability | Portability | Scalability | Correctness | Reusability | Robustness / Fault Tolerance | Complexity / Understandability | Interoperability | Usability | Energy / Memory Efficiency | Security / Safety | Modularity | Testability / Verifiability |
|---|---|---|---|---|---|---|---|---|---|---|---|---|---|---|---|---|
| Implementation | Reuse or Libraries or API | X | | X | X | X | | X | X | X | X | | | | | |
| | Programming Technique | X | X | X | X | X | X | X | | X | | | | | | |
| | Development Framework (propose) | X | X | X | X | X | | | X | | X | X | X | | | |
| | Parallel or Distributed Software | X | X | X | X | X | X | | | | | | | | | |
| | Programming Language | X | X | | | X | | | | X | | | | | X | |
| | Quality Assurance | X | X | | | X | X | X | | X | | | | | | X |
| | Code Generation | X | X | | | | | | | | | | | | | |
| Architecture | CBSE | X | | X | X | X | | X | | | | | | | X | |
| | Design and Architecture Models | X | | X | | | X | X | X | | X | X | X | X | | |
| | Software Architecture | X | | | X | | | X | | | X | | | | | |
| | Parallel or Distributed Software | X | | X | | | X | | | | | | | | | |
| | Domain Specific Language | | | | | | | | | | | | | | | |
| | Programming Technique | | | | | X | X | X | | | | | | | | |

**Figure 8:** Illustrative Example

As an example, suppose a practitioner wants to select the most studied SE practices for improving *maintainability* and *portability* during *implementation* / *architecture*. Based on Figure 8, there are four SE practices (marked with a blue rectangle) that can be used for improving both quality attributes in the aforementioned activities, namely: *Component-Based Software Development*, *Reuse or Libraries or APIs, Programming Techniques*, and *Development of Parallel or Distributed Systems*. However, the practitioner can also use other SE practices for improving *portability* or *maintainability* in isolation (marked with red rectangle). For example, he/she can develop *software architecture models* to address *portability*. Based on the above observation and through a mental qualitative synthesis process, the software engineer understands that the development should be based on reuse: i.e., building component-based architecture, which should be specified through relevant models, and while proceeding at the implementation level he/she should first attempt to reuse code artifacts, such as COTS, third-party libraries etc. Additionally, parallel or distributed architectures should also be considered.

From a research perspective, it is clear that **any research effort must take into account performance**. Even when other quality attributes are of primary interest, the impact of the proposed approach or tool on performance has to be considered. Additionally, we advise researchers not to compromise performance in benefit of other quality attributes. For example, a study focused on the potential benefits of refactoring must also consider the impact on performance. If both effects are present, the researcher should conduct a cost-benefit analysis. The previous statement extends to other quality attributes in the sense that research works, should not only consider one QA, but multiple ones, when proposing an approach or the application of an SE practice. Therefore, we highly advise researchers to **seek for trade-off and cut-off analysis** in their studies, in the sense that any SE is seldomly comes without side-effects, or has a uniform impact across different cases.

Furthermore, one of the most prominent (an unexpected) results of this SMS is the community's focus on reuse. In contrast to traditional software engineering, where reuse is mostly ad-hoc, or highly systematic (e.g., SPLs) (Lambropoulos et al., 2018), in scientific software development reuse relies mostly on library reuse. In particular, it seems that there are many research endeavors to develop for reuse, in many cases with documented impact on qualities of interest (e.g., performance). However, it is not evident that this rich pool of software is visible. Thus, we believe that the development of **a repository of reusable artifacts** (such as Maven repository) would highly benefit the community. Finally, we believe that an **endeavor to catalogue programming techniques**, as well as their benefits and drawbacks would be helpful to the community, in the sense that the current state-of-research suggests an experimentation with a wide range of techniques, with limited reoccurrence. This finding suggests that almost all programming techniques are far from be considered as a state-of-practice in scientific software development.

## 6. Threats to Validity

We organize the threats to validity around the guidelines provided by Ampatzoglou et al. (2019). In Section 6.1, we report threats to validity related to study selection. In Section 6.2, we report threats related to data validity. Finally, in Section 6.3, we report threats related to research validity.

6.1 Study Selection Validity

Study selection validity concerns the early phases of the research, i.e., the search process and the filtering of studies. To guarantee that our search process adequately identified all relevant studies, we used a well-defined protocol to select the primary studies, based on strict guidelines (Kitchenham and Charters, 2007). The identification process consisted of an automated search of the most-known DLs. We used a broad search string that only included keywords and synonyms related to two domains, SE and scientific software development. However, it is possible that the search process returned some candidate primary studies that are related to HPC, but are not to scientific software. However, based on the literature more than 70% of HPC apps are non-computer science apps, but scientific software. By considering that also some of the computer science applications can also be used for scientific purposes (e.g., data science), the percentage of false-positives becomes even lower. We believe that this threat is minor in the sense that the vast majority of applications that are deployed in HPC are science-related (Schmidberger and Brügge, 2012). Additionally, as a means for verification, we have contrasted our dataset to the primary studies of the broader previous secondary study in the literature. In particular, we manually checked and verified that all papers included in the work of Farhoodi et al. (2013) and published in the four DLs of interest are included in our pool of primary studies.

Next, during the inclusion / exclusion phase, it is always possible to accidentally exclude relevant studies. To mitigate this threat, we first extensively discussed the criteria to ensure clarity and a common understanding. Then, we used two authors to conduct this process. These two discussed any potential conflicts. After this process, a third author randomly screening a subset (10%) of the studies chosen for inclusion to verify the choice, without identifying any problems. Furthermore, from our searching space we have excluded grey literature, since the goal of the study focuses on the use of empirical evidence, which are almost never published in grey literature.

We were careful to remove any duplicated results, keeping the most extensive version in our set. Also, our study is not suffering from missing non-English papers and the papers published in a limited number of journals and conferences, since our search process was aiming at a large number of publication venues (including DLs as a whole) all publishing papers only in English. Finally, we were able to access all publications because our institutions provide access to DLs.

6.2 Data Validity

The primary data validity threat is related to *data extraction* bias. The first author extracted and manually recorded all relevant data. Due to the potential for subjectivity in this process, two other authors further inspected and refined the collected data, re-validating them. After this procedure, the results were discussed among the first three authors and they resolved any conflicts.

The next potential threat to data validity is *publication bias*. There are two types of publication bias: (a) bias caused by the fact that primary studies are published by a closed and small circle of researchers; and (b) caused by the tendency of publishing positive rather than negative results (Ampatzoglou et al., 2019). In this study, the first type of publication bias is not present because our broad search identified studies from a large group of researchers. Regarding the second type of publication bias, due to the nature of this study we acknowledge the existence of the threat. Therefore, it is likely that our sample of papers overemphasizes positive results simply because many negative results are not published in the peer-reviewed literature. Given that, readers should take care to examine the results in the papers in case they are interested in a particular SE practice to be applied.

In addition, there are other potential threats to data validity that could affect our study. First, *small sample size* is not of concern because we analyzed close to 3,000 studies. Second, *lack of relationships* is not a threat because our study was not aiming to identify any relationships among data, but only to classify. Third, *low quality of primary studies* is a potential threat because, based on the SMS guidelines (Petersen et al., 2008), we did not perform any type of quality assessment because we did not have an explicit research question related to quality. Therefore, because we counted all studies the same in the analysis, regardless of their quality, it is possible that our counts reflect a biased overall picture. Fourth, *selection of variables to be extracted* is not a threat because the straightforward research questions of our study did not raise any conflicts in the discussions among authors on which variables should be extracted. Fifth, we did not identify issues with the use of *statistical analysis*, in the sense that the nature of our research questions did not require hypothesis testing, but only basic statistical analysis (descriptive statistics). Sixth, to mitigate the *researchers' bias in data interpretation and analysis,* the authors discussed the data clustering for the SE practices and the qualities of interest. Finally, we note that the findings in this mapping study only summarize the state-of-research in this field and not necessarily the state-of-practice. In other words, the study cannot guarantee that the reported results can be generalized as a reflection of industrial practices.

6.3 Research Validity

For the first threat in this category, *research method bias*, the authors are experienced with conducting secondary studies having conducted and reviewed a large number of such studies. Therefore, this threat is minimal. A possible threat to validity of this the selection of the wrong type of research method to answer the research questions. Thus, one concern might be that $RQ_{2.2}$ and $RQ_{2.3}$ have aspects of SLR-questions, because they demand some synthesis. This issue does not threaten the validity of this study, in the sense that in the literature there are various cases (e.g., (Galster et al., 2014) and (Kitchenham et al., 2010)) of hybrid secondary studies—i.e., uses one research method and answers (usually) one RQ, based on the other design. Nevertheless, we note that in our study, the reporting of empirical methods is not done in full detail, i.e., we are only recognizing the type of the empirical method, and not get into details of data collection / analysis methods, samples, etc. which would indeed require more synthesis. For the second threat in this category, *repeatability*, by following a detailed review process we enable the reliability and replication of our study. This manuscript describes the review procedures in detail. Multiple authors were involved in all phases of the process to reduce potential bias. Finally, we have made all extracted data publicly available[7] to enable the comparison and validation of the results.

Additionally, through discussion among the authors we have defined two main research questions that accurately and holistically map to the study goal. This is clearly depicted by the mapping of each research question to the research sub-goals/objectives. Therefore, there was no bias in the selection of specific research questions. Furthermore, in the literature we have been able to identify a substantial amount of related works that can be used for comparison to our results. In particular, for this reason we used related studies from the SE and scientific software development literature. Additionally, the selection of the research method is adequate for the goal of this study and no deviations from the guidelines have been made.

# 7. Summary

Scientists are increasingly turning to the development of software to help them reach their research goals which require the use of large-scale simulations, models, and big data analysis. The size and complexity of these software applications, the need to reuse code for improving productivity, and the need for continuous maintenance have required scientific software developers to become more familiar with SE practices and use them in their projects. This mapping study provided some insight into how and why scientific software developers use SE practices. To obtain as many relevant studies as possible, we search four well-known digital libraries using a well-constructed search string resulting in approximately 1000 articles initially identified. Using a rigorous filtering process, we reduced that total to 359 primary studies, which we analyzed.

---

[7] https://se.uom.gr/wp-content/uploads/SLR_HPC_SE.xlsx

The results showed scientific software development teams are mostly interested in software implementation and testing activities. However, we also found a number of practices that aid in achieving a better architecture and implementation; having a special focus on reuse, either through libraries or components. Also, apart from the understandable focus on performance, maintainability and development productivity stand out as important quality drivers for scientific software developers. To make the results of this study more actionable, we provide an illustration of their usage in a scenario from the perspective of both researchers and practitioners. Finally, we provided researchers of both the SE and scientific software development domains with multiple implications and interesting future work opportunities.

**Acknowledgement**

This work has received funding from the European Union's Horizon 2020 research and innovation programme under grant agreement No 801015 - EXA2PRO (https://exa2pro.eu)